\documentclass[preprint,3p]{elsarticle} 
\usepackage{multicol,caption}

\newenvironment{Figure}
 {\par\medskip\noindent\minipage{\linewidth}}
 {\endminipage\par\medskip}

\usepackage{amssymb}

\usepackage{lineno}

\usepackage{graphicx}
\usepackage{dcolumn}
\usepackage{bm}
\usepackage{epsfig}
\usepackage{rotating}
\usepackage{float}

\journal{Physics Letters B}

\begin{document}

\begin{frontmatter}



\title{Measurement of the $\mathrm e^+\mathrm e^-\rightarrow\mathrm\pi^+\mathrm\pi^-$ Cross Section between 600 and \mbox{900 MeV} Using Initial State Radiation}

\author{
    \begin{center}
      M.~Ablikim$^{1}$, M.~N.~Achasov$^{9,f}$, X.~C.~Ai$^{1}$,
      O.~Albayrak$^{5}$, M.~Albrecht$^{4}$, D.~J.~Ambrose$^{44}$,
      A.~Amoroso$^{49A,49C}$, F.~F.~An$^{1}$, Q.~An$^{46,a}$,
      J.~Z.~Bai$^{1}$
% ...
      \\
      \vspace{0.2cm}
      (BESIII Collaboration)\\
      \vspace{0.2cm}
      {\it
        $^{1}$ Institute of High Energy Physics, Beijing 100049, People's Republic of China\\
% ...
$^{4}$ Bochum Ruhr-University, D-44780 Bochum, Germany\\
        $^{5}$ Carnegie Mellon University, Pittsburgh, Pennsylvania 15213, USA\\
% ...
$^{9}$ G.I. Budker Institute of Nuclear Physics SB RAS (BINP), Novosibirsk 630090, Russia\\
% ...
$^{44}$ University of Rochester, Rochester, New York 14627, USA\\
% ...
$^{46}$ University of Science and Technology of China, Hefei 230026, People's Republic of China\\
% ...
$^{49}$ (A)University of Turin, I-10125, Turin, Italy; (B)University of Eastern Piedmont, I-15121, Alessandria, Italy; (C)INFN, I-10125, Turin, Italy\\
% ...
$^{a}$ Also at State Key Laboratory of Particle Detection and Electronics, Beijing 100049, Hefei 230026, People's Republic of China\\
% ...
$^{f}$ Also at the Novosibirsk State University, Novosibirsk, 630090, Russia\\
% ...
      }\end{center}
    \vspace{0.4cm}
}


\begin{abstract}
We extract the $e^+e^-\rightarrow \pi^+\pi^-$ cross section in the energy range between 600 and 900 MeV, exploiting the method of initial state radiation. A data set with an integrated luminosity of 2.93 fb$^{-1}$ taken at a center-of-mass energy of 3.773 GeV  with the BESIII detector at the BEPCII collider is used. The cross section is measured with a systematic uncertainty of 0.9\%. We extract the pion form factor $|F_\pi|^2$ as well as the contribution of the measured cross section to the leading-order hadronic vacuum polarization contribution to $(g-2)_\mu$. We find this value to be $a_\mu^{\pi\pi,\rm LO}(600-900\;\rm MeV) = (368.2 \pm 2.5_{\rm stat} \pm 3.3_{\rm sys})\cdot 10^{-10}$, which is between the corresponding values using the BaBar or KLOE data.
\end{abstract}


\end{frontmatter}

\begin{multicols}{2}


\section{Introduction}
\label{Motivation}

The cross section $\sigma_{\pi\pi} = \sigma(e^+e^-\rightarrow \pi^+\pi^-)$ has been measured in the past with ever increasing precision at accelerators in Novosibirsk \cite{2pi_CMD2_2004, 2pi_CMD2_2006, 2pi_SND}, Orsay~\cite{2pi_Orsay}, Frascati \cite{2pi_KLOE05,2pi_KLOE08,2pi_KLOE10,2pi_KLOE12}, and SLAC  \cite{2pi_BaBar_PRL,2pi_BaBar}. More recently, the two most precise measurements have been performed by the KLOE collaboration in Frascati \cite{2pi_KLOE12} and the BaBar collaboration at SLAC \cite{2pi_BaBar_PRL,2pi_BaBar}. Both experiments claim a precision of better than 1\% in the energy range below 1 GeV, in which the $\rho(770)$ resonance with its decay into pions dominates the total hadronic cross section. A discrepancy of approximately 3\% on the peak of the $\rho(770)$ resonance is observed between the KLOE and BaBar spectra. The discrepancy is even increasing towards higher energies above the peak of the $\rho$ resonance.
Unfortunately, this discrepancy is limiting the current knowledge of the anomalous magnetic moment of the muon $a_\mu \equiv (g-2)_\mu/2$ \cite{Jegerlehner}, a precision observable of the Standard Model (SM). The accuracy of the  SM prediction of $(g-2)_\mu$ is entirely limited by the knowledge of the hadronic vacuum polarization contribution, which is obtained in a dispersive framework by using experimental data on $\sigma(e^+e^-\rightarrow \rm hadrons)$ \cite{Jegerlehner,g-2_strong_4,Teubner}. The cross section $\sigma(e^+e^- \to \pi^+\pi^-)$ contributes to more than 70\% to this dispersion relation and, hence, is the most important exclusive hadronic channel of the total hadronic cross section. Currently, a discrepancy of 3.6 standard deviations~\cite{g-2_strong_4} is found between the direct measurement of $a_\mu$ and its SM prediction. However, the discrepancy reduces to 2.4$\sigma$ \cite{muonAnomaly_BABARonly}, when only BaBar data is used as input to the dispersion relation. In this letter we present a new measurement of the cross section $\sigma_{\pi\pi}$, obtained by the BESIII experiment at the BEPCII collider in Beijing.

The measurement exploits the method of initial state radiation (ISR), the same method as used by BaBar and KLOE. In the ISR method events are used in which one of the beam particles radiates a high-energy photon. In such a way, the available energy to produce a hadronic (or leptonic) final state is reduced, and the hadronic (or leptonic) mass range below the center-of-mass (cms) energy of the $e^+e^-$ collider becomes available. In this paper, we restrict the studies to the mass range between 600 and \mbox{900 MeV/c$^2$}, which corresponds to the $\rho$ peak region.

The remainder of this letter is organized as follows: In section 2, the BESIII experiment is introduced. In section 3 we describe the data set used, the Monte Carlo (MC) simulation, the event selection of \mbox{$e^+e^-\rightarrow\pi^+\pi^-\gamma$} events, and the data-MC efficiency corrections. The determination of the integrated luminosity of the data set is described in Section~\ref{section_lumi}. A cross check of the used efficiency corrections using the well-known $e^+e^-\rightarrow \mu^+\mu^-\gamma$ QED process is performed in Section \ref{section_QEDtest}, before extracting the $\pi^+\pi^-$ cross section in Section \ref{section_crossSection}.

\section{The BESIII experiment}
The BESIII detector is located at the double-ring Beijing electron-positron collider (BEPCII)~\cite{BESIII}.

The cylindrical BESIII detector covers 93\% of the full solid angle. It consists of the following detector systems.
(1) A Multilayer Drift Chamber (MDC), filled with helium gas, composed of 43 layers, which provides a spatial resolution of 135 $\mu$m, an ionization energy loss $dE/dx$ resolution better than 6\%, and a momentum resolution of 0.5\% for charged tracks at 1~GeV/$c$.
(2) A Time-of-Flight system (TOF), built with 176 plastic scintillator counters in the barrel part, and 96 counters in the endcaps. The time resolution is 80 ps in the barrel and 110 ps in the endcaps. For momenta up to 1 GeV/$c$, this provides a 2$\sigma$ K/$\pi$ separation.
(3) A CsI(Tl) Electro-Magnetic Calorimeter (EMC), with an energy resolution of 2.5\% in the barrel and 5\% in the endcaps at an energy of 1 GeV.
(4) A superconducting magnet producing a magnetic field of 1T.
(5) A Muon Chamber (MUC) consisting of nine barrel and eight endcap resistive plate chamber layers with a 2 cm position resolution.

\section{Data sample, event selection, and efficiency corrections}
\subsection{Data sample and MC simulations}
We analyze 2.93 fb$^{-1}$ (see Sect.~\ref{section_lumi}) of data taken at a cms energy \mbox{$\sqrt{s}$ = 3.773 GeV}, which were collected in two separate runs in 2010 and 2011. The Phokhara event generator \cite{Phokhara,Phokhara7} is used to simulate the signal process $e^+e^-\rightarrow \pi^+\pi^-\gamma$ and the dominant background channel $\mu^+\mu^-\gamma$. The generator includes ISR and final state radiation (FSR) corrections up to next-to-leading order (NLO). Effects of ISR-FSR interference are included as well. The continuum $q\bar q$ ($q=u,d,s$) MC sample is produced with the {\sc kkmc} event generator~\cite{KKMC}. Bhabha scattering events are simulated with \mbox{{\sc babayaga} 3.5~\cite{BABAYAGA}}. The Bhabha process is also used for the luminosity measurement. All MC generators have been interfaced with the \textsc{Geant4}-based detector simulation~\cite{GEANT1,GEANT2}.

\subsection{Event selection}
\label{section_selections}
Events of the type \mbox{$e^+e^-\rightarrow\pi^+\pi^-\gamma$} are selected.
Only a tagged ISR analysis is possible in the mass range  \mbox{600 $<$ $m_{\pi\pi}$ $<$ 900 MeV/c$^2$}, where $m_{\pi\pi}$ is the $\pi^+\pi^-$ invariant mass, {\it i.e.,} the radiated photon has to be explicitly detected in the detector. For untagged events, the photon escapes detection along the beam pipe;  the hadronic system recoiling against the ISR photon is therefore also strongly boosted towards small polar angles, resulting in no geometrical acceptance in the investigated $m_{\pi\pi}$ range.

We require the presence of two charged tracks in the MDC with net charge zero. The points of closest approach to the interaction point (IP) of both tracks have to be within a cylinder with 1 cm radius in the transverse direction and $\pm$10 cm of length along the beam axis. For three-track events, we choose the combination with net charge zero for which the tracks are closest to the IP.
The polar angle $\theta$ of the tracks is required to be found in the fiducial volume of the MDC, $0.4 \,\mathrm{rad} < \theta < \pi - 0.4 \,\mathrm{rad}$, where $\theta$ is the polar angle of the track with respect to the beam axis. We require the transverse momentum $p_t$ to be above 300 MeV/$c$ for each track.
In addition, we require the presence of at least one neutral cluster in the EMC without associated hits in the MDC. We require a deposited energy above 400 MeV. This cluster is then treated as the ISR photon candidate.

The radiative Bhabha process $e^+e^-\rightarrow e^+e^-\gamma(\gamma)$ has a cross section which is up to three orders of magnitude larger than the signal cross section. Electron tracks, therefore, need  to be suppressed. An electron particle identification (PID) algorithm is used for this purpose, exploiting information from the MDC, TOF and EMC \cite{BESIII_2}. The probabilities for being a pion $P(\pi)$ and being an electron $P(e)$ are calculated, and $P(\pi) > P(e)$ is required for both charged tracks.

Using as input the momenta of the two selected track candidates, the energy of the photon candidate, as well as the four-momentum of the initial $e^+e^-$ system, a four-constraint (4C) kinematic fit enforcing energy and momentum conservation is performed which tests the hypothesis $e^+e^-\rightarrow\pi^+\pi^-\gamma$. Events are considered to match the hypothesis if they fulfill the requirement $\chi^2_{4\rm C} < 60$.
It turns out that the $\mu^+\mu^-\gamma$ final state can not be suppressed by means of kinematic fitting due to the limited momentum resolution of the MDC. An independent separation of pion and muon tracks is required.

We utilize a track-based muon-pion separation, which is based on the Artificial Neural Network (ANN) method, as provided by the TMVA package~\cite{TMVA}. The following observables are exploited for the separation: the Zernicke moments of the EMC clusters \cite{BESIII_2}, induced by pion or muon tracks, the ratio of the energy $E$ of a track deposited in the EMC and its momentum $p$ measured in the MDC, the ionization energy loss $dE/dx$ in the MDC, and the depth of a track in the MUC. The ANN is trained using $\pi^+\pi^-\gamma$ and $\mu^+\mu^-\gamma$ MC samples.   We choose the implementation of a Clermont-Ferrand Multilayer Perceptron (CFMlp) ANN as the method resulting in the best background rejection for a given signal efficiency.
The output likelihood $y_{\rm ANN}$ is calculated after training the ANN for the signal pion tracks and background muon tracks.
The response value $y_{\rm ANN}$ is required to be greater than 0.6 for each pion candidate in the event selection, yielding a background rejection of more than 90\% and a signal loss of less than 30\%.

\subsection{Efficiency corrections}
\label{section_effCorr}
Given the accuracy of $\mathcal{O}(1\%)$ targeted for the cross section measurement, possible discrepancies between data and MC due to imperfections of the detector simulation need to be considered. We have investigated data and MC distributions concerning the tracking performance, the energy measurement, and the PID probabilities, both for the electron PID as well as the pion-muon separation. In order to produce test samples of muon and pion tracks over a wide range in momentum/energy and polar angle, we select samples of $\mu^+\mu^-\gamma$ and $\pi^+\pi^-\pi^+\pi^-\gamma$ events that have impurities at the per mille level.  By comparing the efficiencies found in data with the corresponding results found in the MC samples, we determine possible discrepancies. Corresponding correction factors are computed in bins of the track momentum or energy and the track polar angle $\theta$, and are applied to MC tracks to adjust the reconstructed number of events. While for the reconstruction of charged tracks and neutral clusters and for electron PID, the differences between data and MC are smaller than 1\% on average, differences up to 10\% occur in the ANN case.
The corrections are applied separately for neutral clusters and for muon and pion tracks. Hence, we do not only obtain the corrections for the $\pi^+\pi^-\gamma$ signal events, but also for the dominating $\mu^+\mu^-\gamma$ background. The statistical errors of the correction factors are included in the statistical uncertainty of the measurement. Systematic uncertainties associated to the correction factors are presented in Sect.~\ref{section_sys}.  The efficiency correction for the photon efficiency is obtained after the application of the kinematic fit procedure. The corresponding correction is therefore a combined correction of photon efficiency and differences between data and MC of the $\chi^2_{\rm 4C}$ distribution. The systematic uncertainty for the contribution of the photon efficiency and $\chi^2_{\rm 4C}$ distribution is, hence, incorporated in the systematic effects associated with the efficiency corrections. The systematic uncertainty connected with the $p_t$ requirement is also associated with the corresponding efficiency correction.

\subsection{Background subtraction}
The $\mu^+\mu^-\gamma$ background remaining after the application of the ANN is still of the order of a few percent, compared to 5$\times$10$^{5}$ signal events. It is, however, known with high accuracy, as will be shown in the next section, and is subtracted based on MC simulation. Additional background beyond $\mu^+\mu^-\gamma$ remains below the one per mille level. Table~1 lists the remaining MC events after applying all requirements and scaling to the luminosity of the used data set.

\begin{Figure}
	\begin{center}
	\captionof{table}{Total number of remaining non-muon background events between 600 $<$ $m_{\pi\pi}$ $<$ 900 MeV/$c^2$ obtained with MC samples.}
	\begin{tabular}{ c  |  c }
	\hline \hline
	Final state						&	Background events	\\  \hline
	$e^+e^-(n\gamma)$				& $12.0 \pm 3.5$ \\
	$\pi^+\pi^-\pi^0\gamma$			& $3.3\pm 1.8$\\
	$\pi^+\pi^-\pi^0\pi^0\gamma$		& negl.	\\
	$K^+K^-\gamma$				& $2.0\pm 1.5$	\\
	$K^0\overline{ K^0}\gamma$		& $0.4\pm 0.6$\\
	$p\overline{p}\gamma$			& negl.\\
	continuum					& $3.9 \pm 1.9$\\
	$\psi(3770)\rightarrow D^+D^-$					& negl.	\\
	$\psi(3770)\rightarrow D^0\overline{D^0}$			& negl.   \\
	$\psi(3770)\rightarrow \rm non \; D\overline{D}$			& $3.1\pm1.8$\\
	$\gamma \; \psi(2S)$				& negl. \\
	$\gamma \; J/\psi$				& $0.6\pm0.8$ \\
	\hline \hline
	\end{tabular}
	\end{center}
	\label{table_background}
\end{Figure}

\section{Luminosity measurement using Bhabha events}
\label{section_lumi}
The integrated luminosity of the data set used in this work was previously measured in Ref.~\cite{lumi} with a precision of 1.0\% using Bhabha scattering events.
In the course of this analysis, we re-measure the luminosity and decrease its systematic uncertainty by the following means:
(1) Usage of the {\sc babayaga@NLO} \cite{BABAYAGA2} event generator with a theoretical uncertainty of 0.1\%, instead of the previously used {\sc babayaga} 3.5 event generator with an uncertainty of 0.5\% \cite{BABAYAGA}.
(2) Precise estimation of the signal selection efficiencies. In particular, the uncertainty estimate of the polar angle acceptance is evaluated by data-MC studies within the fiducial EMC detection volume, which is relevant for the luminosity study (0.13\%). The very conservative estimate in \cite{lumi} was based on acceptance comparisons with and without using the transition region between the EMC barrel and endcaps, leading to additional data-MC differences (0.75\%).
The other uncertainties of \cite{lumi} remain unchanged and additional systematic uncertainties due to the uncertainty of $\sqrt{s}$ (0.2\%) and the vacuum polarization correction ($<0.01$\%) are taken into account.
Finally, the total integrated luminosity amounts to $\mathcal L =(2931.8\pm0.2_{\rm stat}\pm13.8_{\rm sys})$pb$^{-1}$ with a relative uncertainty of 0.5\%, which is consistent with the previous measurement~\cite{lumi}.

\section{QED test using $e^+e^-\rightarrow\mu^+\mu^-\gamma$ events}
\label{section_QEDtest}
The yield of events of the channel \mbox{$e^+e^-\rightarrow\mu^+\mu^-\gamma$} as a function of the two-muon invariant mass $m_{\mu\mu}$ can be compared to a precise prediction by QED, which is provided by the Phokhara generator. We select muon events according to the ANN method described previously and require $y_{\rm ANN} < 0.4$ for both tracks, resulting in a background rejection of more than 90\% and a signal loss of less than 20\%. All other requirements in the selection are exactly the same as for the $\pi^+\pi^-\gamma$ analysis. The remaining pion background after the $\mu^+\mu^-\gamma$ selection is much reduced, reaching 10\% in the $\rho$ peak region. A comparison between data and MC is shown in Fig.~\ref{QED_test}.
The same data sample as used in the main analysis is also used here, but we present a larger mass range than for the $\pi^+\pi^-\gamma$ case. The efficiency corrections described in the previous section have been applied to MC on a track and photon candidate basis. The lower panel of Fig.~\ref{QED_test} shows the relative discrepancy between data and MC. A good agreement over the full $m_{\mu\mu}$ mass range at the level of  ($1.0\pm 0.3\pm0.9$)\% and $\chi^2/{\rm ndf} = 134/139$ is found, where the uncertainties are statistical and systematic, respectively. A difference in the mass resolution due to detector effects between data and MC is visible around the narrow $J/\psi$ resonance. A fit in the mass range 600 $<m_{\mu\mu}<$ 900 MeV/$c^2$, which is the mass range studied in the main analysis, gives a relative discrepancy of (2.0 $\pm$ 1.7 $\pm$ 0.9)\%; this is illustrated in the inset of the upper panel of Fig.~\ref{QED_test}. The theoretical uncertainty of the MC generator Phokhara is below 0.5\% \cite{Phokhara}, while the
systematic uncertainty of our measurement is 0.9\%. The latter is dominated by the luminosity measurement, which is needed for the normalization of the data set.
We consider the good agreement between the $\mu^+\mu^-\gamma$ QED prediction and data as a validation of the accuracy of our efficiency corrections. As a further cross check, we have applied the efficiency corrections also to a statistically independent $\mu^+\mu^-\gamma$ sample, resulting in a difference between data and MC of ($0.7\pm 0.2$)\% over the full mass range, where the error is statistical only.

\begin{Figure}
   \centering
    \includegraphics[trim = 50 2 3 4,clip,width=8.9cm]{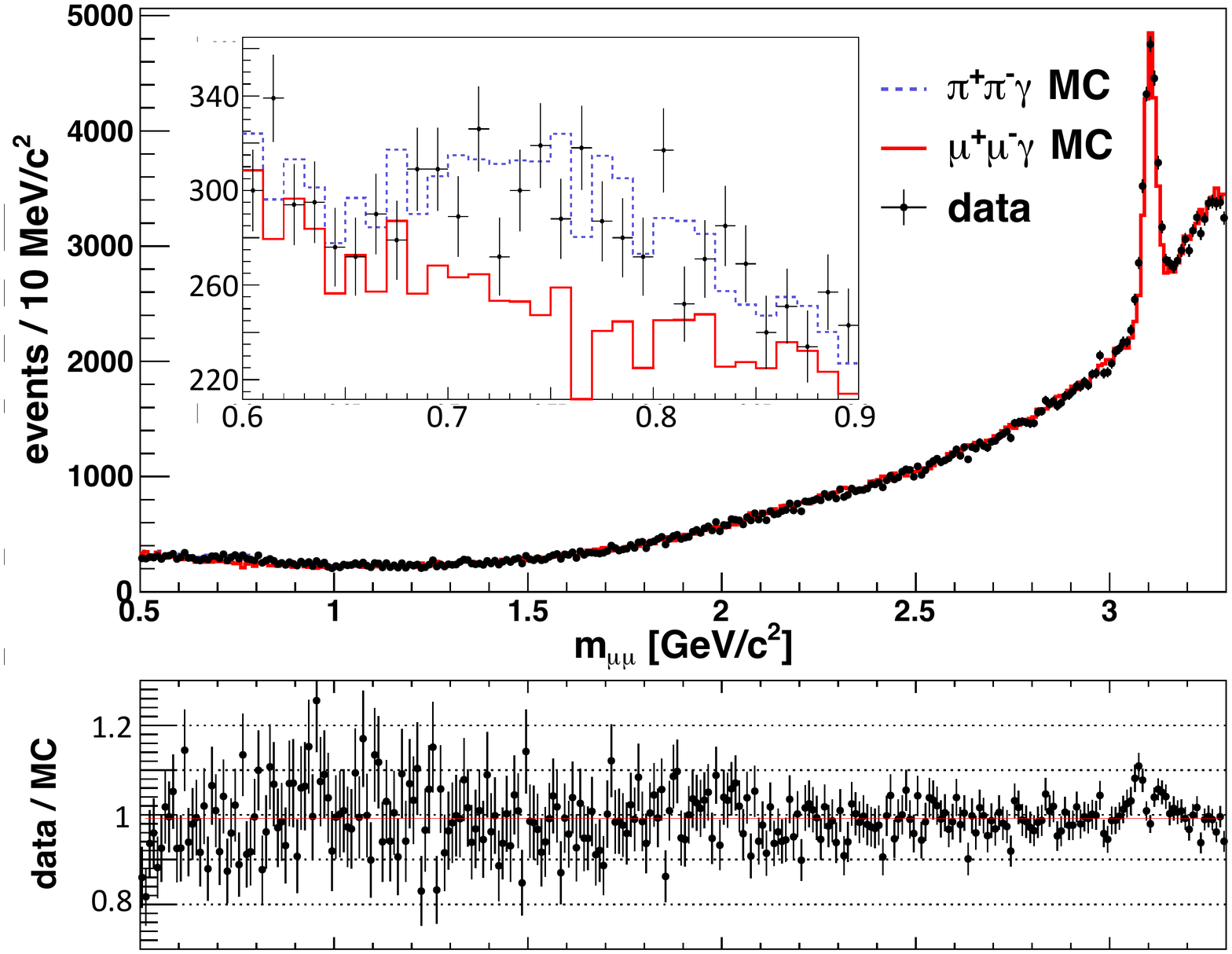}
     \captionof{figure}{Invariant $\mu^+\mu^-$ mass spectrum of data and $\mu^+\mu^-\gamma$ MC after using the ANN as muon selector and applying the efficiency corrections. The upper panel presents the absolute comparison of the number of events found in data and MC. The inset shows the zoom for invariant masses between 0.6 and 0.9 GeV/$c^2$. The MC sample is scaled to the luminosity of the data set. The lower plot shows the ratio of these two histograms. A linear fit is performed to quantify the data-MC difference, which gives a difference of \mbox{(1.0 $\pm$ 0.3 $\pm$ 0.9)\%}. A difference in the mass resolution between data and MC is visible around the narrow $J/\psi$ resonance.}
   \label{QED_test}
\end{Figure}

\section{Extraction of $\sigma(e^+e^-\rightarrow\pi^+\pi^-)$ and $|F_\pi^2|$}
\label{section_crossSection}

\subsection{Methods}
We finally extract $\sigma_{\pi\pi} = \sigma(e^+e^-\rightarrow \pi^+\pi^-)$ according to two independent normalization schemes. In the first method, we obtain the bare cross section, {\it i.e.,} the cross section corrected for vacuum polarization effects, according to the following formula:
\begin{equation}
\label{formula_preferred}
	\sigma_{\pi\pi(\gamma_{\rm FSR})}^{\rm bare} = \frac{N_{\pi\pi\gamma}\cdot (1 + \delta_{\rm FSR}^{\pi\pi}) }{\mathcal L \cdot \epsilon_{\rm global}^{\pi\pi\gamma} \cdot H(s) \cdot \delta_{\rm vac}} \; ,
\end{equation}
where $N_{\pi\pi\gamma}$ is the number of signal events found in data after applying all selection requirements described above and an unfolding procedure to correct for the mass resolution, $\mathcal L$ the luminosity of the data set, and $H$ the radiator function. The global efficiency $\epsilon_{\rm global}^{\pi\pi\gamma}$ is determined based on the signal MC by dividing the measured number of events after all selection requirements $N^{\rm true}_{\rm measured}$ by that of all generated events $N^{\rm true}_{\rm generated}$. The true MC sample is used, with the full $\theta_\gamma$ range, applying the efficiency corrections mentioned in Section \ref{section_effCorr} but without taking into account the detector resolution in the invariant mass $m$:
\begin{equation}
\label{equ_efficiency}
	\epsilon_{\rm global}(m)  = \frac{N^{\rm true}_{\rm measured}(m) }{N^{\rm true}_{\rm generated}(m) } \; .
\end{equation}
The efficiency is found to depend slightly on $m_{\pi\pi}$ and ranges from 2.8\% to 3.0\% from lowest to highest $m_{\pi\pi}$. An unfolding procedure, which eliminates the effect of the detector resolution, is described in Sect.~\ref{section_unfolding} and is applied before dividing by the global efficiency. The radiator function $H$ is described in Sect.~\ref{section_vacPol}. As input for $a_\mu$ the bare cross section is needed. It can be obtained by dividing the cross section by  the vacuum polarization correction $\delta_{\rm vac}$, which is also described in Sect.~\ref{section_vacPol}. As pointed out in Ref.~\cite{Jegerlehner}, in order to consider radiative effects in the dispersion integral for $a_\mu$, an FSR correction has to be performed.
The determination of the correction factor $(1 + \delta_{\rm FSR}^{\pi\pi})$ is described in Sect.~\ref{section_FSRcorr}.

In the second method, we use a different normalization than in the first method and normalize $N_{\pi\pi\gamma}$ to the measured number of $\mu^+\mu^-\gamma$ events, $N_{\mu\mu\gamma}$. Since $\mathcal L$, $H$, and $\delta_{\rm vac}$ cancel in this normalization, one finds the following formula:
\begin{equation}
\label{formula_normalize}
	\sigma_{\pi\pi(\gamma_{\rm FSR})}^{\rm bare} = \frac{N_{\pi\pi\gamma} }{N_{\mu\mu\gamma} } \cdot\frac{\epsilon_{\rm global}^{\mu\mu\gamma} }{\epsilon_{\rm global}^{\pi\pi\gamma}}\cdot \frac{1+ \delta_{\rm FSR}^{\mu\mu} }{1 + \delta_{\rm FSR}^{\pi\pi} } \cdot \sigma_{\mu\mu}^{\rm bare}  \; ,
\end{equation}
where $\epsilon_{\rm global}^{\mu\mu\gamma}$ is the global efficiency of the dimuon selection, already described in Sect.~\ref{section_QEDtest}, $\delta_{\rm FSR}^{\mu\mu}$ is the FSR correction factor to the $\mu^+\mu^-$ final state, which can be obtained using the Phokhara event generator,
$\sigma_{\mu\mu}^{\rm bare}$ is the exact QED prediction of the dimuon cross section, given by \cite[Eq. (5.13)]{sigma_muon}
\begin{equation}
\label{equ_dimuon}
\sigma_{\mu\mu}^{\rm bare}  = \frac{4\pi\alpha^2}{3s'}\cdot\frac{\beta_\mu(3-\beta_\mu^2)}{2} \; ,
\end{equation}
with the fine structure constant $\alpha$, the cms energy $s' < s$ available for the creation of the final state, the muon velocity $\beta_\mu = \sqrt{1 - 4m_{\mu}^2/s'}$, and the muon mass $m_\mu$.
The contributions of radiator function, luminosity, and vacuum polarization to the systematic uncertainties of the bare cross section, cancel in the second method. The upper panel of Fig.~\ref{crossCheck_normalize} shows the comparison of the bare cross sections including FSR obtained with the first (black) and second method before unfolding (blue). The error bars are statistical only. They are much larger for the second method due to the limited $\mu^+\mu^-\gamma$ statistics in the mass range of interest. The lower panel shows the ratio of these cross sections. Again, a linear fit is performed to quantify the difference, which is found to be (0.85 $\pm$ 1.68)\% and $\chi^2/{\rm ndf} = 50/60$, where the error is statistical. Both methods agree within uncertainties. The first one is used in the analysis.
Finally, the pion form factor as a function of $s'$ can be calculated via
\begin{equation}
	|F_\pi|^2 (s') = \frac{3s'}{\pi\alpha^2\beta_{\pi}^3(s')}\sigma_{\pi\pi}^{\rm dressed}(s') \; ,
\end{equation}
with the pion velocity $\beta_\pi(s') = \sqrt{1 - 4m_{\pi}^2/s'}$, the charged pion mass $m_\pi$, and the dressed cross section $\sigma_{\pi\pi}^{\rm dressed}(s') = \sigma(e^+e^-\rightarrow \pi^+\pi^-)(s')$ containing vacuum polarization, but corrected for FSR effects. The result is presented in Sect.~\ref{section_results}.

\begin{Figure}
   \centering
   \includegraphics[width=8.5cm]{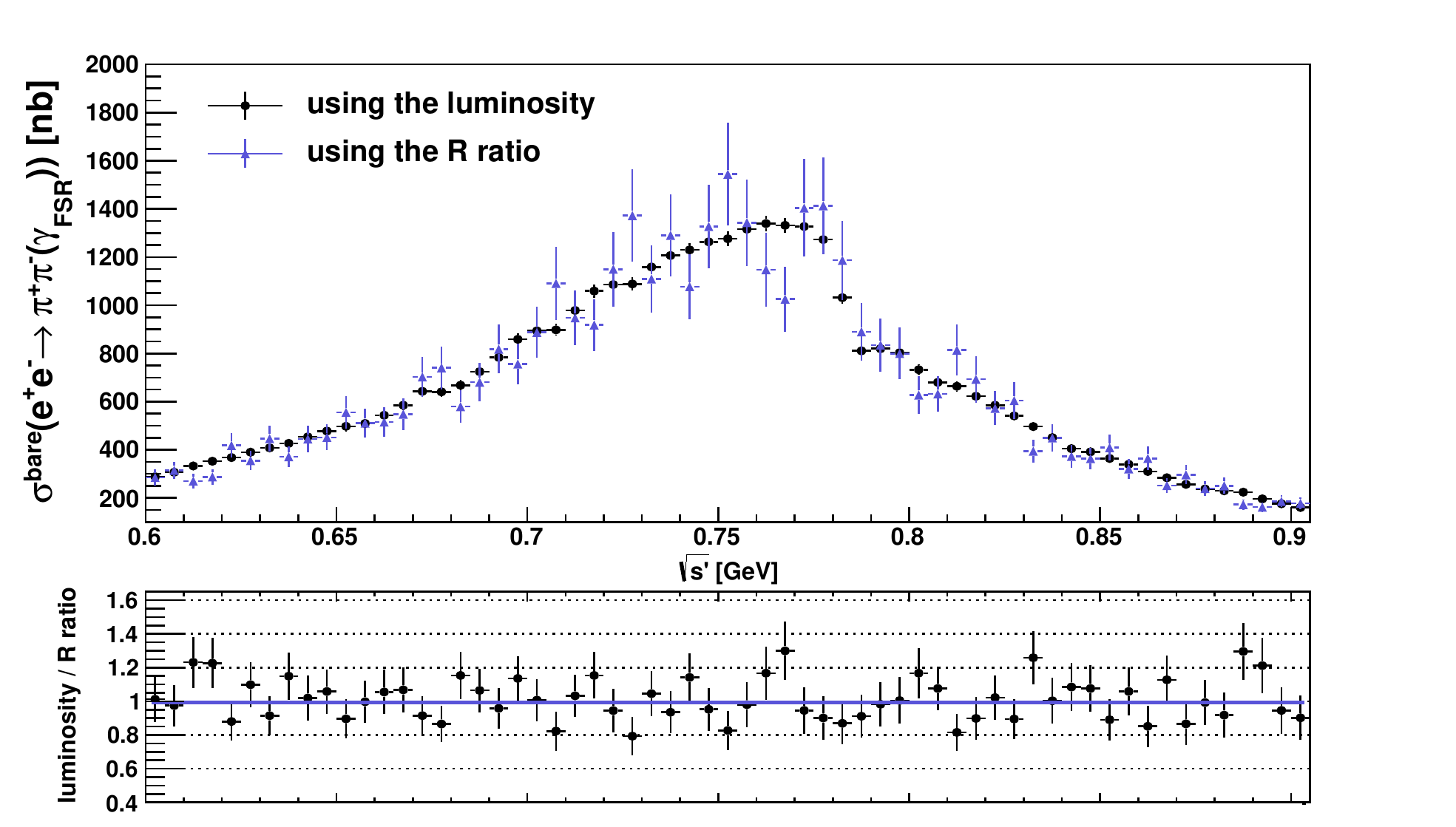}
     \captionof{figure}{Comparison between the methods to extract $\sigma_{\pi\pi}$ explained in the text --- using the luminosity (black) and normalizing by $\sigma_{\mu\mu}$ (blue). The lower panel shows the ratio of these results together with a linear fit (blue line) to quantify their difference.}
   \label{crossCheck_normalize}
\end{Figure}

\subsection{Unfolding}
\label{section_unfolding}
In order to obtain the final result for $\sigma_{\pi\pi}$, one has to rectify the detector resolution effects, {\it i.e.,} the mass spectrum needs to be unfolded. To this end, the Singular Value Decomposition (SVD) method \cite{unfolding} is used. It requires two input variables --- the response matrix and the regularization parameter $\tau$. The SVD algorithm calculates an operator which cancels the detector smearing by inverting the response matrix.
We obtain the response matrix in the full mass range between threshold and 3.0 GeV, using a signal MC sample. The matrix corresponds to the correlation of the reconstructed $m_{\pi\pi}$ spectrum, and the originally generated $m_{\pi\pi}$ values.
With the choice of a bin width of 5 MeV/$c^2$, about 43\% of events are found to be on the diagonal axis.

To find the value of the regularization parameter $\tau$, we compare two independent methods, as suggested in Ref.~\cite{unfolding}. On the one hand, we perform a MC simulation where $\tau$ is optimized such that unfolded and true distributions have the best agreement. On the other hand, we process an algorithm, described in \cite{unfolding}, exploiting the singular values of the response matrix. Both methods favor a similar regularization parameter of $\tau \cong 72$.

To estimate the systematic uncertainties and to test the stability of the SVD method, we perform two cross checks. In both cases we use a $\pi^+\pi^-\gamma$ MC sample which is independent of the one used to determine the response matrix. We modify and then unfold the spectra in both checks. In the first cross check, the reconstructed spectrum is smeared with an additional Gaussian error, which results in an about 20\% larger detector smearing than expected from MC simulation. The resulting unfolded spectrum reproduces the true one on the sub per mille level. In the second cross check, the mass of the $\rho$-resonance is varied systematically in the simulation in steps of 10 MeV/$c^2$ between 750 and 790 MeV/$c^2$. The response matrix is kept fixed and was determined with a $\rho$ mass of 770 MeV/$c^2$. In all cases, the masses of the $\rho$ peak after unfolding are found to be close to the initially simulated masses.
From the comparisons of these checks, we take the maximum deviation of 0.2\% as systematic uncertainty.

\subsection{FSR correction}
\label{section_FSRcorr}
The correction factor $\delta_{\rm FSR}$ is determined with the Phokhara generator in bins of $m_{\pi\pi}$. Two different correction methods are used on the data to cross check whether it is applied correctly.

(1) The whole FSR contribution of the $\pi^+\pi^-\gamma$ events is calculated with Phokhara, by dividing a true MC spectrum including FSR in NLO by the spectrum without any FSR contribution. The resulting distribution is used to correct data.
As pointed out in Ref.~\cite{Jegerlehner}, for the dispersion integral for $a_\mu$, the FSR correction for the process $e^+e^-\rightarrow\pi^+\pi^-$ needs then to be added again. We use the calculation by Schwinger assuming point-like pions:
\begin{equation}
	\sigma_{\pi\pi(\gamma)}^{\rm dressed} = \sigma_{\pi\pi}^{\rm dressed}\cdot \bigg[1 + \eta(s)\frac{\alpha}{\pi}\bigg] \; ,
\end{equation}
where $\eta(s)$ is the theoretical correction factor taken from \cite{FSR_Schwinger}. In the $\rho$-peak region it is between 0.4\% and 0.9\%.

(2) A special version of the Phokhara generator is used \cite{Phokhara_omega}, which, in contrast to the standard version of the generator, distinguishes whether a photon is emitted in the initial or the final state. In events in which photons have been radiated solely due to ISR, the momentum transfer of the virtual photon $s_{\gamma^*}$ is equal to the invariant mass of the two pions $m_{\pi\pi}^2$. However, if an FSR photon is emitted, the invariant mass is lowered due to this effect and hence $m_{\pi\pi}^2 < s_{\gamma^*}$. The effect can be removed by applying an unfolding procedure, using again the SVD algorithm. Here, the response matrix is $m_{\pi\pi}^2$ vs. $s_{\gamma^*}$, obtained from a MC sample that includes FSR in NLO. The regularization parameter $\tau$ is determined as described in Sect.~\ref{section_unfolding}. After applying the corrections for the radiative $\pi^+\pi^-\gamma$ process, which are of the order of 2\%, one obtains the $\pi^+\pi^-(\gamma_{\rm FSR})$ cross section directly.

The difference between both methods is found to be \mbox{(0.18 $\pm$ 0.13)\%}. Both methods are complementary and agree with each other within errors. The difference is taken as systematic uncertainty. Finally, the correction obtained with method (1) is used in the analysis.

\subsection{Radiator function and vacuum polarisation correction}
\label{section_vacPol}
The radiator function is implemented within the Phokhara event generator with NLO precision. Hence, a very precise description is available with a claimed uncertainty of 0.5\%~\cite{Phokhara}.

To obtain the \emph{bare} cross section, vacuum polarization effects $\delta_{\rm vac}$ must be taken into account. To this aim, the dressed cross section, including the vacuum polarization effects, is adjusted for the running of the coupling constant $\alpha$ \cite{vacPol}. Bare and dressed cross sections are related as follows:
\begin{equation}
	\sigma^{\rm bare} = \frac{\sigma^{\rm dressed}}{\delta_{\rm vac}} = \sigma^{\rm dressed}\cdot\bigg( \frac{\alpha(0)}{\alpha(s)}\bigg)^2 \; .
\end{equation}
The correction factors are taken from Ref.~\cite{vacPol_link}.

\subsection{Summary of systematic uncertainties}
\label{section_sys}
Systematic uncertainties are studied within the investigated $m_{\pi\pi}$ range between 600 and 900~MeV/$c^2$. Sources are: \\
(1) Efficiency corrections: Each individual uncertainty is studied in bins of $m_{\pi\pi}$ with respect to three different sources. Firstly, the remaining background contaminations in the data samples are estimated with the corresponding MC simulation mentioned in Tab.~1. Their contribution is taken into account by multiplying the claimed uncertainties of the event generators and their fraction of the investigated signal events.
Secondly, we vary the selection requirements ($E/p$, $\chi^2_{1C}$, depth of a charged track in the MUC), which are used to select clean muon and pion samples for the efficiency studies, in a range of three times the resolution of the corresponding variable. The differences of the correction factors are calculated. Thirdly, the resolution of the correction factors, {\it i.e.,} the bin sizes of momentum and $\theta$ distributions, is varied by a factor two and the effects on the final correction factors are tested. \\
(2) Pion-muon separation: Additional uncertainties of using the ANN method  for pion-muon separation are estimated by comparing the result from a different multivariate method, namely the Boosted Decision Tree (BDT) approach \cite{TMVA}.  As a further cross check, the whole analysis is repeated without the use of a dedicated PID method. \\
(3) Residual background is subtracted using simulated events. The uncertainty is determined to be 0.1\%. \\
(4) Angular acceptance: The knowledge of the angular acceptance of the tracks is studied by varying this requirement by more than three standard deviations of the angular resolution and  studying the corresponding difference in the selected number of events. A difference of 0.1\% in the result can be observed. The procedure is repeated for all other selection criteria. Their contribution to the total systematic uncertainty is found to be negligible. \\
(5) Unfolding: Uncertainties introduced by unfolding are smaller than 0.2\%, as estimated by the two cross checks mentioned in Sect.~\ref{section_unfolding}.\\
(6) FSR correction: The uncertainty due to the FSR correction is obtained by comparing two different approaches as described in Sect.~\ref{section_FSRcorr}. The uncertainty is found to be 0.2\%. \\
(7) Vacuum Polarization: The uncertainty due to the vacuum polarization correction is conservatively estimated to be 0.2\%. \\
(8) Radiator Function: The Radiator Function extracted from the Phokhara generator is implemented with a precision of 0.5\%. \\
(9) Luminosity: The luminosity of the analyzed data set has been determined to a precision of 0.5\%.\\
All systematic uncertainties are summarized in Tab.~\ref{systematic_summary}. They are added in quadrature, and a total systematic uncertainty for $\sigma^{\rm bare}(e^+e^-\rightarrow\pi^+\pi^-(\gamma_{\rm FSR}))$ of 0.9\% is achieved, which is fully correlated amongst all
data points.
\begin{Figure}
	\centering
	\captionof{table}{Summary of systematic uncertainties.}
	\begin{tabular}{ c | c }
	\hline \hline
	Source	&	 Uncertainty    \\
			&	(\%)\\
	\hline
	Photon efficiency correction		&	0.2	\\
	Pion tracking efficiency correction	&	0.3	\\
	Pion ANN efficiency correction		&	0.2	\\
	Pion e-PID efficiency	 correction	&	0.2	\\
	ANN				 			& 	negl. \\
	Angular acceptance				&	0.1  	\\
	Background subtraction			&	0.1	\\
	Unfolding						&	0.2	\\
	FSR correction	$\delta_{\rm FSR}$		&	0.2	\\
	Vacuum polarization correction $\delta_{\rm vac}$	&	0.2 	\\
	Radiator function				&	0.5	\\
	Luminosity $\mathcal L$			&	0.5	\\	\hline
	\bf Sum						&	\bf 0.9	\\
	\hline \hline
	\end{tabular}
	\label{systematic_summary}
\end{Figure}

\section{Results}
\label{section_results}
The result for $\sigma^{\rm bare}(e^+e^-\rightarrow\pi^+\pi^-(\gamma_{\rm FSR}))$ as a function of $\sqrt{s}=m_{\pi\pi}$ is illustrated in Fig.~\ref{result_crossSection} and given numerically in Tab.~\ref{table_sigma}. The cross section is corrected for vacuum polarization effects and includes final state radiation. Besides the dominant $\rho(770)$ peak, the well-known structure of the $\rho$-$\omega$ interference is observed. The result for the pion form factor $|F_\pi|^2$ is shown in Fig.~\ref{result_FF} and given numerically in Tab.~\ref{table_sigma}. It includes vacuum polarization corrections, but, differently from the cross section shown in Fig.~\ref{result_crossSection}, final state radiation effects are excluded here. The red line in Fig.~\ref{result_FF} illustrates a fit to data

\begin{Figure}
	\centering
	\captionof{table}{Fit parameters and statistical errors of the Gounaris-Sakurai fit of the pion form factor. Also shown are the PDG 2014 values \cite{PDG2014}.}
	\begin{tabular}{ l  c c c}
	\hline \hline
	Parameter		&	BESIII value 	&	PDG 2014\\
	\hline
	$m_\rho$ [MeV/$c^2$]		&	776.0 $\pm$ 0.4 	&	775.26 $\pm$ 0.25\\
	$\Gamma_\rho$ [MeV]		&	151.7 $\pm$ 0.7	&	147.8 $\pm$ 0.9 \\
	$m_\omega$ [MeV/$c^2$]	&	782.2 $\pm$ 0.6 	&	782.65 $\pm$ 0.12\\
	$\Gamma_\omega$ [MeV]	&	fixed to PDG 	&	8.49   $\pm$ 0.08\\
	$|$c$_\omega|$ [10$^{-3}$]	&	1.7     $\pm$ 0.2	&	-\\
	$|\phi_\omega|$ [rad]		&	0.04  $\pm$ 0.13	&	-\\
	\hline \hline
	\end{tabular}
	\label{table_fit}
\end{Figure}

\begin{Figure}
   \centering
   \includegraphics[width=8.5cm]{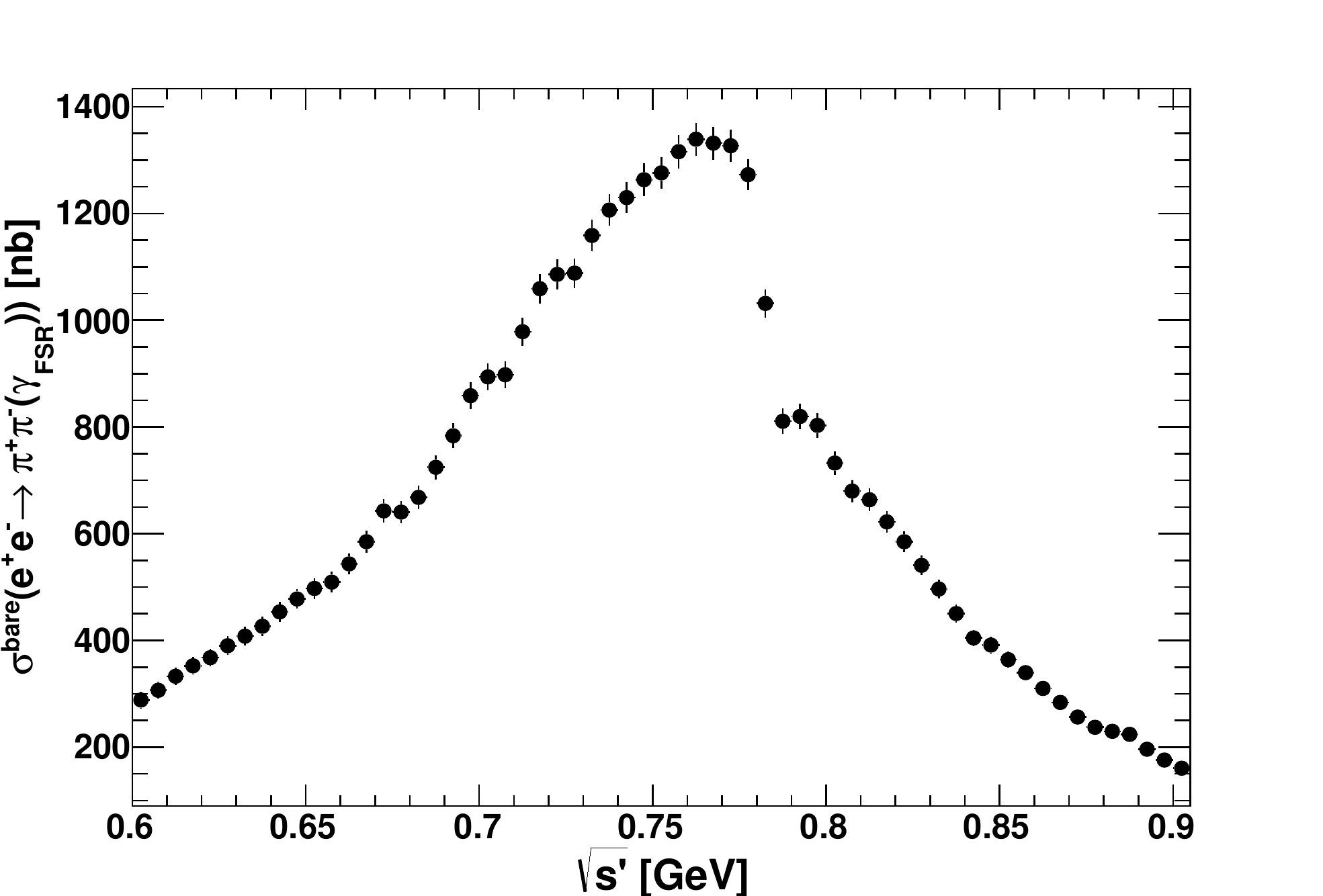}
     \captionof{figure}{The measured bare $e^+e^-\rightarrow\pi^+\pi^-(\gamma_{\rm FSR})$ cross section. Only the statistical errors are shown.}
   \label{result_crossSection}
\end{Figure}

\begin{Figure}
   \centering
   \includegraphics[width=8.5cm]{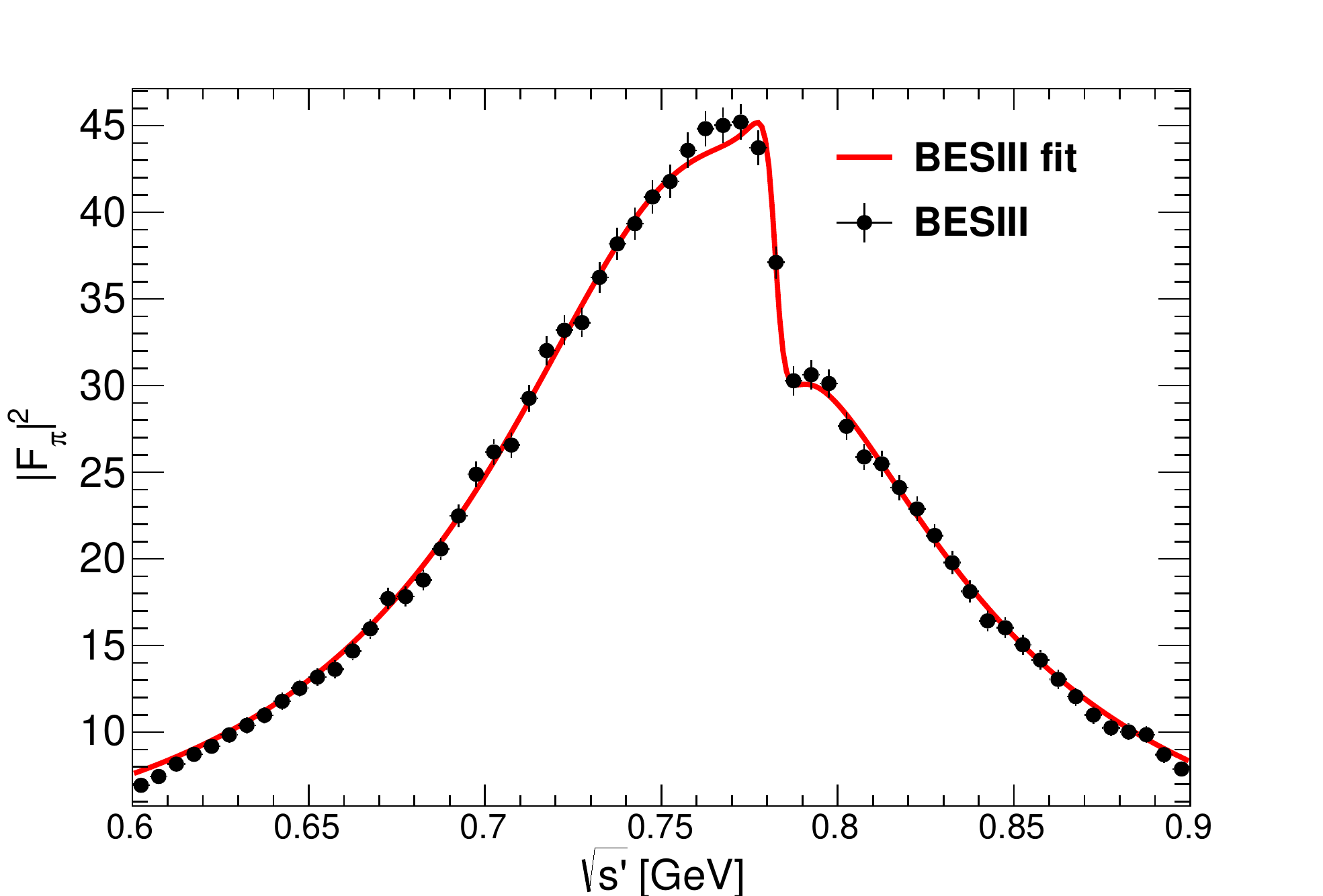}
     \captionof{figure}{The measured squared pion form factor $|F_{\pi}|^2$. Only statistical errors are shown. The red line represents the fit using the Gounaris-Sakurai parametrization.}
   \label{result_FF}
\end{Figure}

\noindent according to a parametrization proposed by Gounaris and Sakurai \cite{GS}. Here, exactly the same fit formula and fit procedure are applied as described in detail in Ref.~\cite{2pi_BaBar}. Free parameters of the fit are the mass and width $\Gamma$ of the $\rho$ meson, the mass of the $\omega$ meson, and the phase of the Breit-Wigner function $c_\omega = |c_\omega| e^{i\phi_\omega}$. The width of the $\omega$ meson is fixed to the PDG value \cite{PDG2014}. The resulting values are shown in Tab.~\ref{table_fit}. As can be seen, the resonance parameters are in agreement with the PDG values \cite{PDG2014} within uncertainties, except for $\Gamma_{\rho}$, which shows a 3.4$\sigma$ deviation. Corresponding amplitudes for the higher $\rho$ states, $\rho(1450)$, $\rho(1700)$, and $\rho(2150)$, as well as the masses and widths of those states were taken from Ref.~\cite{2pi_BaBar}, and the systematic uncertainty in $\Gamma_\rho$ due to these assumptions has not been quantitatively evaluated.

\begin{figure*}[t!]
\begin{minipage}{\columnwidth}
\begin{Figure}
   \centering
   \includegraphics[width=8.5cm]{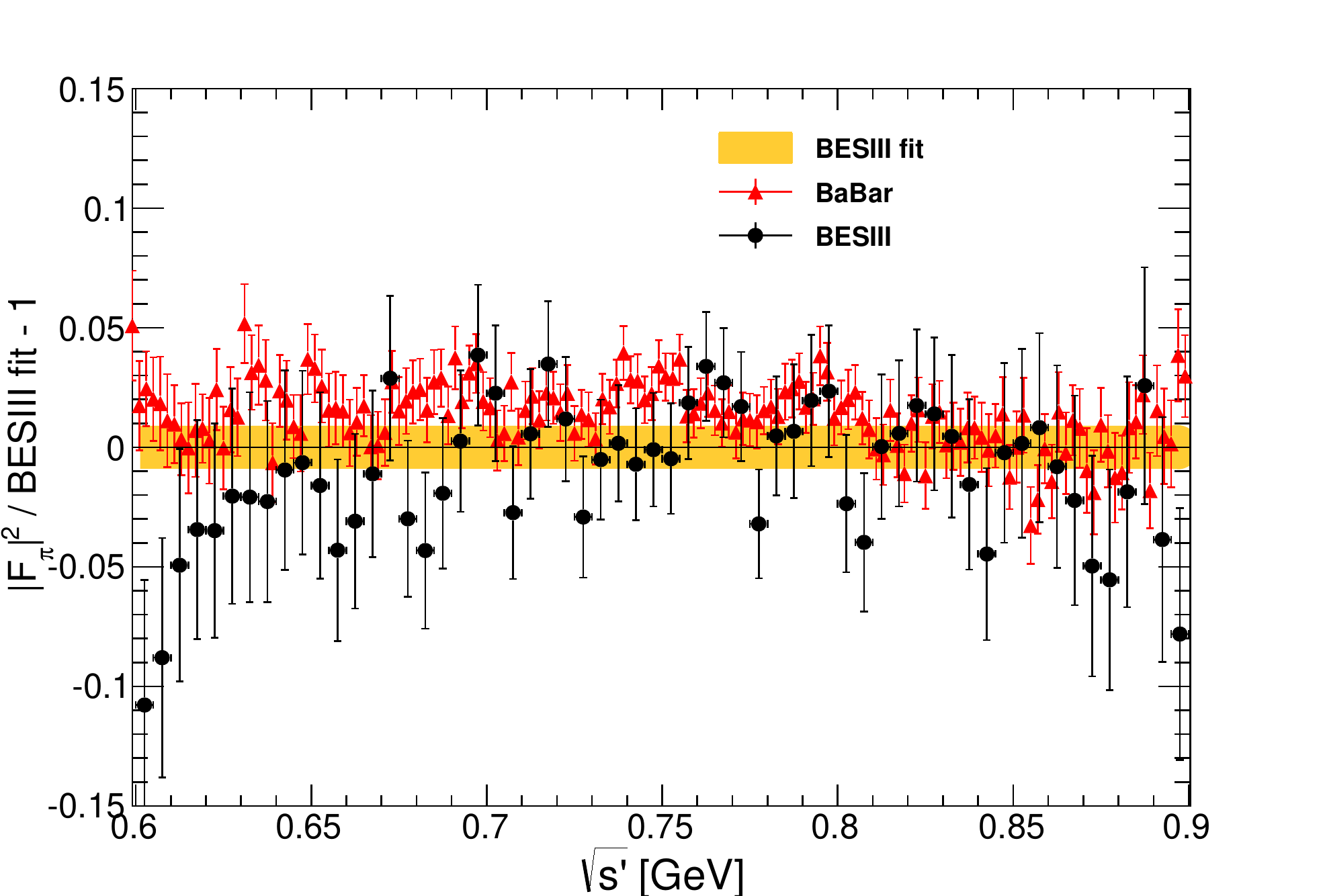}
     \captionof{figure}{Relative difference of the form factor squared from BaBar \cite{2pi_BaBar} and the BESIII fit. Statistical and systematic uncertainties are included in the data points. The width of the BESIII band shows the systematic uncertainty only.}
   \label{result_FF_Babar}
\end{Figure}
\end{minipage}
\hfill
\begin{minipage}{\columnwidth}
\begin{Figure}
   \centering
   \includegraphics[width=8.5cm]{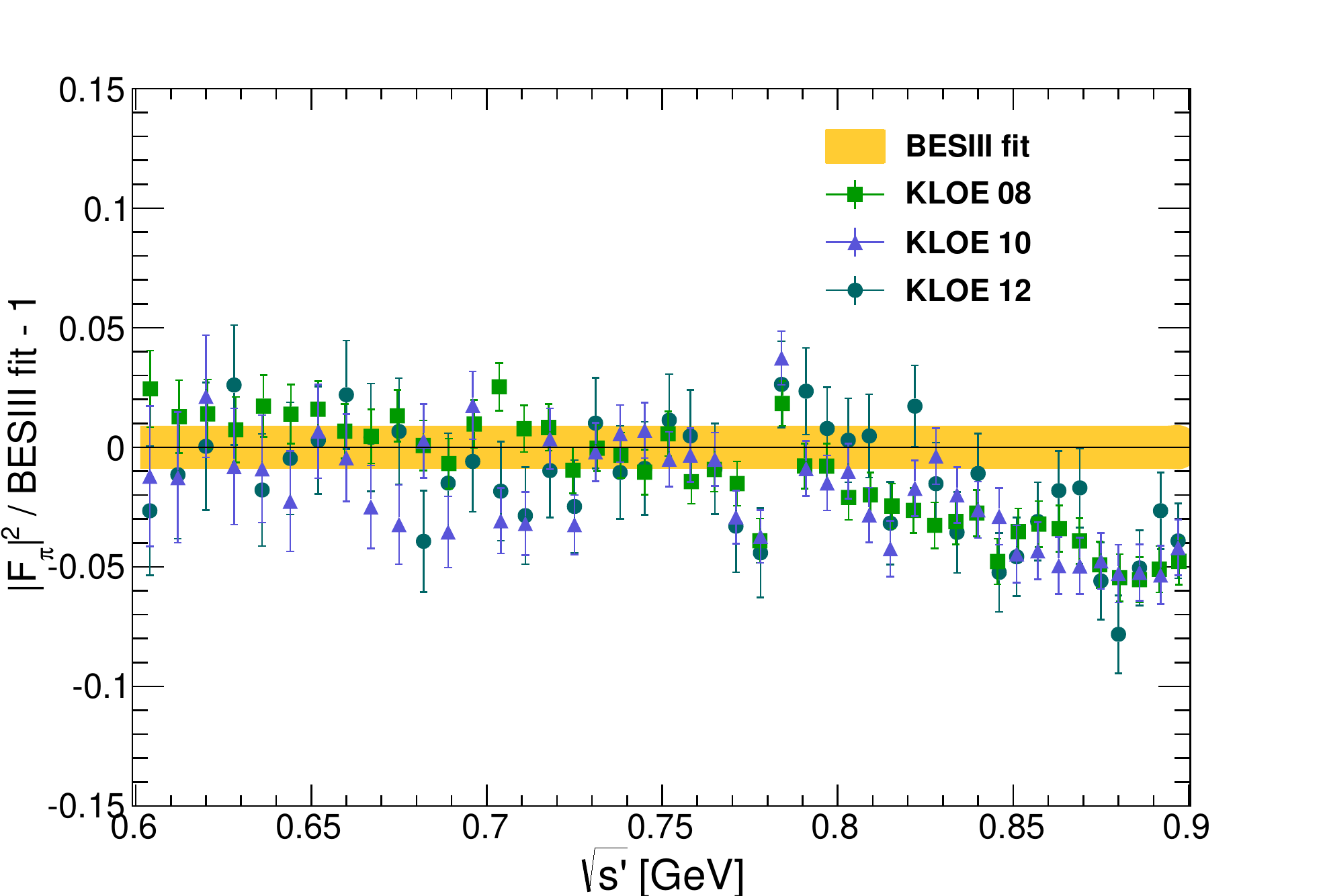}
     \captionof{figure}{Relative difference of the form factor squared from KLOE  \cite{2pi_KLOE08,2pi_KLOE10,2pi_KLOE12} and the BESIII fit. Statistical and systematic uncertainties are included in the data points. The width of the BESIII band shows the systematic uncertainty only.}
   \label{result_FF_KLOE}
\end{Figure}
\end{minipage}
\end{figure*}

The Gounaris-Sakurai fit provides an excellent description of the BESIII data in the full
mass range from 600 to 900 MeV/$c^2$, resulting in $\chi^2/{\rm ndf} = 49.1/56$. Figure \ref{result_FF_Babar} shows the difference between fit and data. Here the data points show the statistical uncertainties only, while the shaded error band of the fit shows the systematic uncertainty only.

In order to compare the result with previous measurements, the relative
difference of the BESIII fit and data from BaBar \cite{2pi_BaBar},  KLOE \cite{2pi_KLOE08,2pi_KLOE10,2pi_KLOE12}, CMD2~\cite{2pi_CMD2_2004,2pi_CMD2_2006}, and SND \cite{2pi_SND} is investigated. Such a comparison is complicated by the fact, that previous measurements used different vacuum polarization corrections. Therefore, we consistently used the vacuum polarization correction from Ref.~\cite{vacPol_link} for all the comparisons discussed in this section. The KLOE 08, 10, 12, and BaBar spectra have, hence, been modified accordingly. The individual comparisons are illustrated in Figs.~\ref{result_FF_Babar} and \ref{result_FF_KLOE}. Here, the shaded error band of the fit includes the systematic error only, while the uncertainties of the data points include the sum of the statistical and systematic errors.
We observe a very good agreement with the KLOE 08 and KLOE 12 data sets up to the mass range of the $\rho$-$\omega$ interference. In the same mass range the BaBar and KLOE 10 data sets show a systematic shift, however, the deviation is, not exceeding 1 to 2 standard deviations.
At higher masses, the statistical error bars in the case of BESIII are relatively large, such that a comparison is not conclusive. There seem to be a good agreement with the BaBar data, while a large deviation with all three KLOE data sets is visible. There are indications that the BESIII data and BESIII fit show some disagreement in the low mass and very high mass tails as well.
We have also compared our results in the $\rho$ peak region with data from Novosibirsk. At lower and higher masses, the statistical
uncertainties of the Novosibirsk results are too large to draw definite
conclusions. The spectra from SND and from the 2006 publication of CMD-2 are found to
be in very good agreement with BESIII in the $\rho$ peak region, while the 2004 result of CMD-2 shows a systematic deviation of a few percent.

\begin{figure*}
   \centering
   \includegraphics[width=14cm]{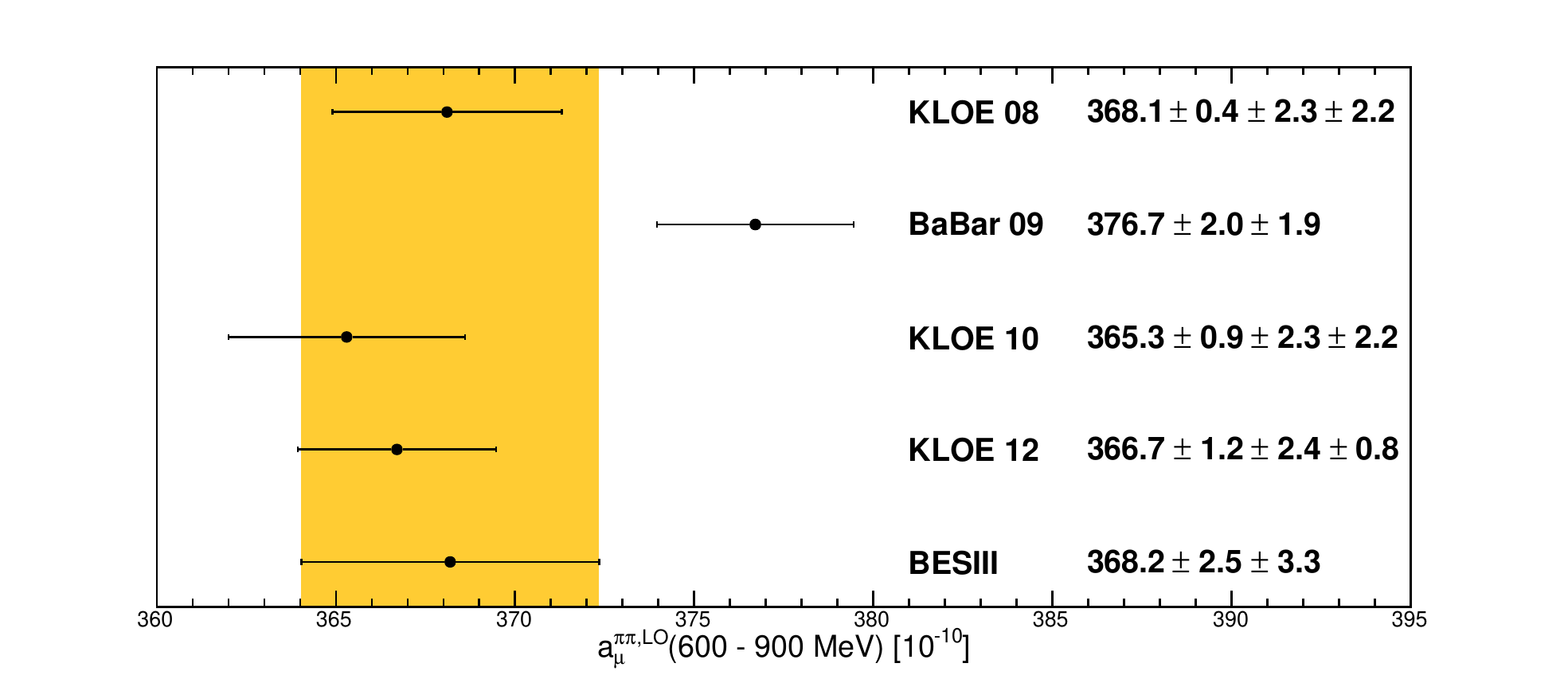}
     \captionof{figure}{Our calculation of the leading-order (LO) hadronic vacuum polarization 2$\pi$ contributions to $(g-2)_\mu$ in the energy range 600 - 900 MeV from BESIII and
based on the data from KLOE 08 \cite{2pi_KLOE08}, 10 \cite{2pi_KLOE10}, 12 \cite{2pi_KLOE12}, and BaBar \cite{2pi_BaBar}, with the statistical and systematic errors. The statistical and systematic errors are added quadratically. The band shows the 1$\sigma$ range of the BESIII result.}
   \label{result_a_mu}
\end{figure*}

We also compute the contribution of our BESIII cross section measurement \mbox{$\sigma^{\rm bare}(e^+e^-\rightarrow\pi^+\pi^-(\gamma_{\rm FSR}))$} to the hadronic contribution of $(g-2)_\mu$,
\small
\begin{equation}
	a_\mu^{\pi\pi,\rm LO}(0.6 - 0.9\, \rm{GeV}) = \frac{1}{4\pi^3}\int\limits_{(0.6\rm{GeV})^2}^{(0.9\rm{GeV})^2} ds' K(s') \sigma^{\rm bare}_{\pi\pi(\gamma)} \; ,
\end{equation}
\normalsize where $K(s')$ is the kernel function \cite[Eq. (5)]{Jegerlehner}.
As summarized in Fig.~\ref{result_a_mu}, the BESIII result, $a_\mu^{\pi\pi,\rm LO}(600-900\;\rm MeV) = (368.2 \pm 2.5_{\rm stat} \pm 3.3_{\rm sys})\cdot 10^{-10}$, is found to be in good agreement with all three KLOE values. A difference of about 1.7$\sigma$ with respect to the BaBar result is observed.

\begin{table*}[ht]
	\centering
	\captionof{table}{Results of the BESIII measurement of the cross section $\sigma^{\rm bare}_{\pi^+\pi^-(\gamma_{\rm FSR})} \equiv \sigma^{\rm bare}(e^+e^-\rightarrow\pi^+\pi^-(\gamma_{\rm FSR}))$ and the squared pion form factor $|F_\pi|^2$. The errors are statistical only. The value of $\sqrt{s'}$ represents the bin center. The 0.9\%
systematic uncertainty is fully correlated between any two bins.}

	\begin{tabular}{ c | c | c || c | c | c}
	\hline \hline
	&&&&& \\
	$\sqrt{s'}$ [MeV]	&	 $\sigma^{\rm bare}_{\pi^+\pi^-(\gamma_{\rm FSR})}$ [nb]		&	$|F_{\pi}|^2$   & $\sqrt{s'}$ [MeV]	&	 $\sigma^{\rm bare}_{\pi^+\pi^-(\gamma_{\rm FSR})}$ [nb]		&	$|F_{\pi}|^2$   \\
	&&&&& \\
	\hline
	&&&&& \\
602.5 & 288.3 $\pm$ 15.2 & 6.9 $\pm$ 0.4 & 
752.5 & 1276.1 $\pm$ 29.8 & 41.8 $\pm$ 1.0 \\ 
607.5 & 306.6 $\pm$ 15.5 & 7.4 $\pm$ 0.4 & 
757.5 & 1315.9 $\pm$ 31.3 & 43.6 $\pm$ 1.0 \\ 
612.5 & 332.8 $\pm$ 16.3 & 8.2 $\pm$ 0.4 & 
762.5 & 1339.3 $\pm$ 30.9 & 44.8 $\pm$ 1.0 \\ 
617.5 & 352.5 $\pm$ 16.3 & 8.7 $\pm$ 0.4 & 
767.5 & 1331.9 $\pm$ 30.8 & 45.0 $\pm$ 1.0 \\ 
622.5 & 367.7 $\pm$ 16.6 & 9.2 $\pm$ 0.4 & 
772.5 & 1327.0 $\pm$ 30.6 & 45.2 $\pm$ 1.0 \\ 
627.5 & 390.1 $\pm$ 17.7 & 9.8 $\pm$ 0.4 & 
777.5 & 1272.7 $\pm$ 29.2 & 43.7 $\pm$ 1.0 \\ 
632.5 & 408.0 $\pm$ 18.0 & 10.4 $\pm$ 0.5 & 
782.5 & 1031.5 $\pm$ 26.7 & 37.1 $\pm$ 0.9 \\ 
637.5 & 426.6 $\pm$ 18.1 & 11.0 $\pm$ 0.5 & 
787.5 & 810.7 $\pm$ 24.2 & 30.3 $\pm$ 0.8 \\ 
642.5 & 453.5 $\pm$ 19.0 & 11.8 $\pm$ 0.5 & 
792.5 & 819.7 $\pm$ 23.8 & 30.6 $\pm$ 0.8 \\ 
647.5 & 477.7 $\pm$ 18.5 & 12.5 $\pm$ 0.5 & 
797.5 & 803.1 $\pm$ 23.3 & 30.1 $\pm$ 0.8 \\ 
652.5 & 497.4 $\pm$ 19.5 & 13.2 $\pm$ 0.5 & 
802.5 & 732.4 $\pm$ 22.1 & 27.7 $\pm$ 0.8 \\ 
657.5 & 509.2 $\pm$ 19.4 & 13.6 $\pm$ 0.5 & 
807.5 & 679.9 $\pm$ 20.6 & 25.9 $\pm$ 0.7 \\ 
662.5 & 543.4 $\pm$ 19.9 & 14.7 $\pm$ 0.5 & 
812.5 & 663.6 $\pm$ 21.0 & 25.5 $\pm$ 0.8 \\ 
667.5 & 585.0 $\pm$ 20.5 & 16.0 $\pm$ 0.6 & 
817.5 & 622.2 $\pm$ 19.9 & 24.1 $\pm$ 0.7 \\ 
672.5 & 642.7 $\pm$ 22.2 & 17.7 $\pm$ 0.6 & 
822.5 & 585.0 $\pm$ 19.5 & 22.9 $\pm$ 0.7 \\ 
677.5 & 640.5 $\pm$ 21.0 & 17.8 $\pm$ 0.6 & 
827.5 & 540.8 $\pm$ 18.1 & 21.4 $\pm$ 0.7 \\ 
682.5 & 668.0 $\pm$ 21.9 & 18.8 $\pm$ 0.6 & 
832.5 & 496.4 $\pm$ 17.7 & 19.8 $\pm$ 0.7 \\ 
687.5 & 724.4 $\pm$ 22.9 & 20.6 $\pm$ 0.6 & 
837.5 & 450.4 $\pm$ 16.8 & 18.1 $\pm$ 0.6 \\ 
692.5 & 783.5 $\pm$ 23.2 & 22.5 $\pm$ 0.7 & 
842.5 & 404.7 $\pm$ 15.2 & 16.4 $\pm$ 0.6 \\ 
697.5 & 858.6 $\pm$ 25.3 & 24.9 $\pm$ 0.7 & 
847.5 & 391.3 $\pm$ 15.4 & 16.0 $\pm$ 0.6 \\ 
702.5 & 893.8 $\pm$ 25.4 & 26.2 $\pm$ 0.7 & 
852.5 & 364.0 $\pm$ 15.0 & 15.0 $\pm$ 0.6 \\ 
707.5 & 897.8 $\pm$ 25.0 & 26.6 $\pm$ 0.7 & 
857.5 & 339.6 $\pm$ 14.0 & 14.2 $\pm$ 0.6 \\ 
712.5 & 978.6 $\pm$ 26.6 & 29.3 $\pm$ 0.8 & 
862.5 & 310.0 $\pm$ 13.7 & 13.0 $\pm$ 0.6 \\ 
717.5 & 1059.1 $\pm$ 27.9 & 32.0 $\pm$ 0.8 & 
867.5 & 283.8 $\pm$ 13.0 & 12.1 $\pm$ 0.5 \\ 
722.5 & 1086.0 $\pm$ 28.3 & 33.2 $\pm$ 0.9 & 
872.5 & 256.5 $\pm$ 12.4 & 11.0 $\pm$ 0.5 \\ 
727.5 & 1088.4 $\pm$ 27.7 & 33.6 $\pm$ 0.9 & 
877.5 & 237.3 $\pm$ 11.4 & 10.3 $\pm$ 0.5 \\ 
732.5 & 1158.8 $\pm$ 29.2 & 36.2 $\pm$ 0.9 & 
882.5 & 229.7 $\pm$ 11.6 & 10.0 $\pm$ 0.5 \\ 
737.5 & 1206.5 $\pm$ 29.6 & 38.2 $\pm$ 0.9 & 
887.5 & 224.0 $\pm$ 11.6 & 9.9 $\pm$ 0.5 \\ 
742.5 & 1229.9 $\pm$ 29.0 & 39.3 $\pm$ 0.9 & 
892.5 & 196.1 $\pm$ 10.5 & 8.7 $\pm$ 0.4 \\ 
747.5 & 1263.3 $\pm$ 30.3 & 40.9 $\pm$ 1.0 & 
897.5 & 175.9 $\pm$ 9.7 & 7.9 $\pm$ 0.4 \\ 
	&&&&& \\
	\hline \hline
	\end{tabular}
	\label{table_sigma}
\end{table*}

\section{Conclusion}
A new measurement of the cross section $\sigma^{\rm bare}(e^+e^-\rightarrow{\pi^+\pi^-(\gamma_{\rm FSR})})$ has been performed with an accuracy of 0.9\% in the dominant $\rho(770)$ mass region between 600 and 900 MeV/$c^{2}$, using the ISR method at BESIII. The energy dependence of the cross section appears compatible with corresponding measurements from KLOE and BaBar within approximately one standard deviation. The two-pion contribution to the hadronic vacuum polarization contribution to $(g-2)_\mu$  has been determined from the BESIII data to be $a_\mu^{\pi\pi,\rm LO}(600-900\;\rm MeV) = (368.2 \pm 2.5_{\rm stat} \pm 3.3_{\rm sys})\cdot 10^{-10}$. By averaging the KLOE, BaBar, and BESIII values of $a_\mu^{\pi\pi,\rm LO}$ and assuming that the five data sets are independent, a deviation of more than 3$\sigma$ between the SM prediction of $(g-2)_\mu$ and its direct measurement is confirmed. For the low mass region $<$ 600 MeV/$c^2$ and the high mass region $>$ 900 MeV/$c^2$, the BaBar data was used in this calculation.

\section{Acknowledgements}
The BESIII collaboration thanks the staff of BEPCII and the IHEP computing center for their strong support. We thank Thomas Teubner for the recalculation of $a_\mu^{\pi\pi,\rm LO}(600-900\;\rm MeV)$ and Fedor Ignatov for the useful discussions. This work is supported in part by National Key Basic Research Program of China under Contract No. 2015CB856700; National Natural Science Foundation of China (NSFC) under Contracts Nos. 11125525, 11235011, 11322544, 11335008, 11425524; the Chinese Academy of Sciences (CAS) Large-Scale Scientific Facility Program; the CAS Center for Excellence in Particle Physics (CCEPP); the Collaborative Innovation Center for Particles and Interactions (CICPI); Joint Large-Scale Scientific Facility Funds of the NSFC and CAS under Contracts Nos. 11179007, U1232201, U1332201; CAS under Contracts Nos. KJCX2-YW-N29, KJCX2-YW-N45; 100 Talents Program of CAS; National 1000 Talents Program of China; INPAC and Shanghai Key Laboratory for Particle Physics and Cosmology; German Research Foundation DFG under Contract No. Collaborative Research Center CRC-1044; Istituto Nazionale di Fisica Nucleare, Italy; Ministry of Development of Turkey under Contract No. DPT2006K-120470; Russian Foundation for Basic Research under Contract No. 14-07-91152; The Swedish Resarch Council; U. S. Department of Energy under Contracts Nos. DE-FG02-04ER41291, DE-FG02-05ER41374, DE-FG02-94ER40823, DESC0010118; U.S. National Science Foundation; University of Groningen (RuG) and the Helmholtzzentrum fuer Schwerionenforschung GmbH (GSI), Darmstadt; WCU Program of National Research Foundation of Korea under Contract No. R32-2008-000-10155-0.





\section*{References}
\begin{thebibliography}{00}

	\bibitem{2pi_CMD2_2004} R.R. Akhmetshin et al. [CMD2 Collaboration], Phys.
Lett. B~$\bf{578}$, 285 (2004).
	\bibitem{2pi_CMD2_2006} R.R. Akhmetshin et al. [CMD2 Collaboration], Phys. Lett. B~$\bf{648}$, 28 (2007).
	\bibitem{2pi_SND} M.N. Achasov et al. [SND Collaboration], JETP~$\bf{101}$, 1053 (2005).

	
	\bibitem{2pi_Orsay} J.E. Augustin et al., Phys.~Rev.~Lett.~$\bf{20}$, 126 (1968).

	\bibitem{2pi_KLOE05} A. Aloisio et al. [KLOE Collaboration], Phys. Lett. B~$\bf 606$, 12 (2005).
	\bibitem{2pi_KLOE08} F. Ambrosio et al. [KLOE Collaboration], Phys. Lett. B~$\bf{670}$, 285 (2009).
	\bibitem{2pi_KLOE10} F. Ambrosino et al. [KLOE Collaboration],
Phys. Lett.~B~$\bf{700}$, 102 (2011).
	\bibitem{2pi_KLOE12} D. Babusci et al. [KLOE Collaboration], Phys. Lett. B~$\bf{720}$, 336 (2013).
	\bibitem{2pi_BaBar_PRL} B. Aubert et al. [BABAR Collaboration], Phys. Rev. Lett.~$\bf{103}$, 231801 (2009).
	\bibitem{2pi_BaBar} J. P. Lees et al. [BABAR Collaboration], Phys. Rev. D~$\bf{86}$, 032013 (2012).


	\bibitem{Jegerlehner} S. Eidelman, F. Jegerlehner, Z.Phys. C~$\bf{67}$, 585 (1995).
	\bibitem{g-2_strong_4} M. Davier, A. Hoecker, B. Malescu and Z. Zhang, Eur. Phys. J. C~$\bf{71}$, 1515 (2011).
	\bibitem{Teubner} K. Hagiwara, R. Liao, A.D. Martin, Daisuke Nomura, T. Teubner, J. Phys. G~$\bf{38}$, 085003 (2011).
	\bibitem{muonAnomaly_BABARonly} M. Davier, A. Hoecker, B. Malescu, C. Z. Yuan and  Z. Zhang, Eur. Phys. J. C~$\bf{66}$, 1 (2010).

	\bibitem{BESIII} M. Ablikim et al. [BESIII Collaboration], Nucl. Instrum. Meth.~A~$\bf{614}$, 345 (2010).



	\bibitem{Phokhara} G. Rodrigo, H. Czy\.{z}, J. H. Kuhn, M. Szopa, Eur.~Phys.~J.~C~$\bf{24}$, 71 (2002).
	\bibitem{Phokhara7} H. Czy\.{z}, J. H. Kuhn and A. Wapienik, Phys. Rev. D~$\bf{77}$, 114005 (2008).

	\bibitem{KKMC} S. Jadach, B. F. L. Ward and Z. Was, Comput. Phys. Commun.~$\bf{130}$, 260 (2000).
	\bibitem{BABAYAGA} G. Balossini, C. M. C. Calame, G. Montagna, O. Nicrosini and F. Piccinini, Nucl. Phys. B~$\bf{758}$, 227 (2006).

	\bibitem{GEANT1} J. Allison et al. [GEANT4 Collaboration], IEEE Transactions on Nuclear Science~$\bf{53}$, 270 (2006).
	\bibitem{GEANT2} S. Agostinelli et al. [GEANT4 Collaboration], Nucl. Instrum. Meth.~A~$\bf{506}$, 250 (2003).
	
	\bibitem{BESIII_2} D. M. Asner et al., Int.~J.~Mod.~Phys.~A~$\bf{24}$, 1 (2009).

	\bibitem{TMVA} A. Hoecker. P. Speckmayer, J. Stelzer, J. Therhaag, E. Von Toerne and H. Voss, PoS ACAT~$\bf{040}$ (2007).


	\bibitem{lumi} M. Ablikim et al. [BESIII Collaboration], Chin. Phys. C~$\bf{37}$, 123001 (2013).
	\bibitem{BABAYAGA2} G. Balossini, C. Bignamini, C.M. Carloni Calame, G. Montagna, F. Piccinini and O. Nicrosini, Phys. Lett. B~$\bf{663}$, 209 (2008).
	
	
	\bibitem{sigma_muon} M. E. Peskin and D. V. Schroeder, \begin{it}An Introduction to Quantum Field Theory\end{it}, Vol. 2, USA, Addison-Wesley, 135 (1995).

	\bibitem{unfolding} A. Hoecker, V. Kartvelishvili, Nucl. Instrum. Meth. A~$\bf{372}$, 469 (1996).

	\bibitem{FSR_Schwinger} J. S. Schwinger, \begin{it}Particles, Sources and Fields\end{it}, Vol. 3, Redwood City, USA Addison-Wesley, 99 (1989).

	\bibitem{Phokhara_omega} Private communication with H. Czyz.

	\bibitem{vacPol} A. H\"ofer, J. Gluza and F. Jegerlehner, Eur. Phys. J. C~$\bf{24}$, 51 (2002).

	\bibitem{vacPol_link} F. Jegerlehner, Nucl. Phys. Proc. Suppl. $\bf{181-182}$, 135 (2008); F. Jegerlehner, Z. Phys. C~$\bf{32}$, 195 (1986);
	\mbox{www-com.physik.hu-berlin.de/$\sim$fjeger/alphaQED.tar.gz} (2015)

	\bibitem{GS} G. J. Gounaris and J. J. Sakurai, Phys. Rev. Lett.~$\bf{21}$,
244 (1968).

	\bibitem{PDG2014} K.A. Olive et al. [Particle Data Group], Chin. Phys. C~$\bf{38}$, 090001 (2014).



\end{thebibliography}




\end{multicols}

\end{document}